\documentclass[12pt]{article}
\usepackage{psfig,epsf}
\usepackage{a4wide,graphicx}
\usepackage{cite}

\begin{document}
\begin{center}
{\Large{\bf {Erratum: Neutron and proton spectra from the decay of
$\Lambda$ hypernuclei [Phys. Rev. C55, 735 (1997)]} }}
\end{center}
\begin{center}
{A. Ramos$^{1}$ , M.J. Vicente-Vacas$^{2}$  and E. Oset$^{2}$ }
\end{center}

\begin{center}

{\small{$^{1}$ \it Departament d'Estructura i Constituents de la Mat\`eria, \\
Universitat de Barcelona, Diagonal 647, E-08028 Barcelona, Spain}}

{\small{$^{2}$ \it Departamento de F\'{\i}sica Te\'orica and IFIC, \\
Centro Mixto Universidad de Valencia-CSIC, \\
Ap. Correos 22085, E-46071 Valencia, Spain}}

\end{center}

\vspace{0.5cm}
 PACS number(s): 21.80.+a, 24.10.Lx
\vspace{0.5cm}

An error in the computer program which evaluated nucleon spectra
following the decay of $\Lambda$ hypernuclei has been detected.
The error affected the final state interaction (FSI) of the
nucleons, reducing by a factor ten the nucleon-nucleon (NN)
collision probability. When correcting the error, some results and
conclusions of Ref. [1] have changed substantially.

In this erratum we present the corrected results, keeping the same
numbering of the original figures to avoid confusion.  The
corrections affect figures 6 to 8 and 11 to 16, which we display
below.

In summary, the effect of FSI of the nucleons is much more
important  than in the old results. This has as consequences,

\begin{itemize}

\item The shape of the spectra with FSI is drastically modified. 
Due to the larger NN collision probability,
more nucleons are removed from the high
energy part of the spectrum and fill up the low energy part.
This can be seen in Fig. 7.

\item The numbers of emitted neutron and protons tend to become more similar
because of the increased number of collisions,
particularly at low energies in the spectrum where nucleons which
have undergone several collisions accumulate. This is clearly
visualized in Fig. 11, which shows the ratio $N_n/N_p$ for
different experimental cuts.

\item The number of protons ($N_p$) or neutrons ($N_n$) per decay event is
strongly dependent on the energy cut, as can be seen in Figs. 12,
13 and 14. For instance, removing the protons with energies
smaller than 30 MeV reduces $N_p$ by a factor about three in
$^{12}_\Lambda$C and more than four in $^{56}_\Lambda$Fe.

\item The sensitivity of the number of protons or neutrons per decay
event to $\Gamma_n / \Gamma_p$ is weaker than before. For
instance, whereas the number of emitted protons in
$^{12}_\Lambda$C increased by 45 percent when $\Gamma_n /
\Gamma_p$ changed from 0 to 2,  now the increase is only 33
percent. One of the consequences is that, given an experimental
value for the number of neutrons or protons with a certain error,
the uncertainties induced in the $\Gamma_n / \Gamma_p$ ratio will
be somewhat larger than before.

\item The agreement with the shape of the experimental spectra is now much better
than before for a wide range of values of $\Gamma_n / \Gamma_p$,
as one can see in Figs. 15 and 16. From the spectrum of Fig. 15,
low values of $\Gamma_n / \Gamma_p$ around or below unity seem to
be preferred. The situation is less clear for the spectrum of Fig.
16 which has larger error bars.

\end{itemize}

   The new results presented here might affect some experimental analyses
   which have used the results of \cite{Ramos:1996ik} as input.

\section*{Acknowledgments}
We would like to thank G. Garbarino for pointing out this error to us.

\setcounter{figure}{5}

\begin{figure}[htb]
\centerline{
     \includegraphics[width=0.8\textwidth]{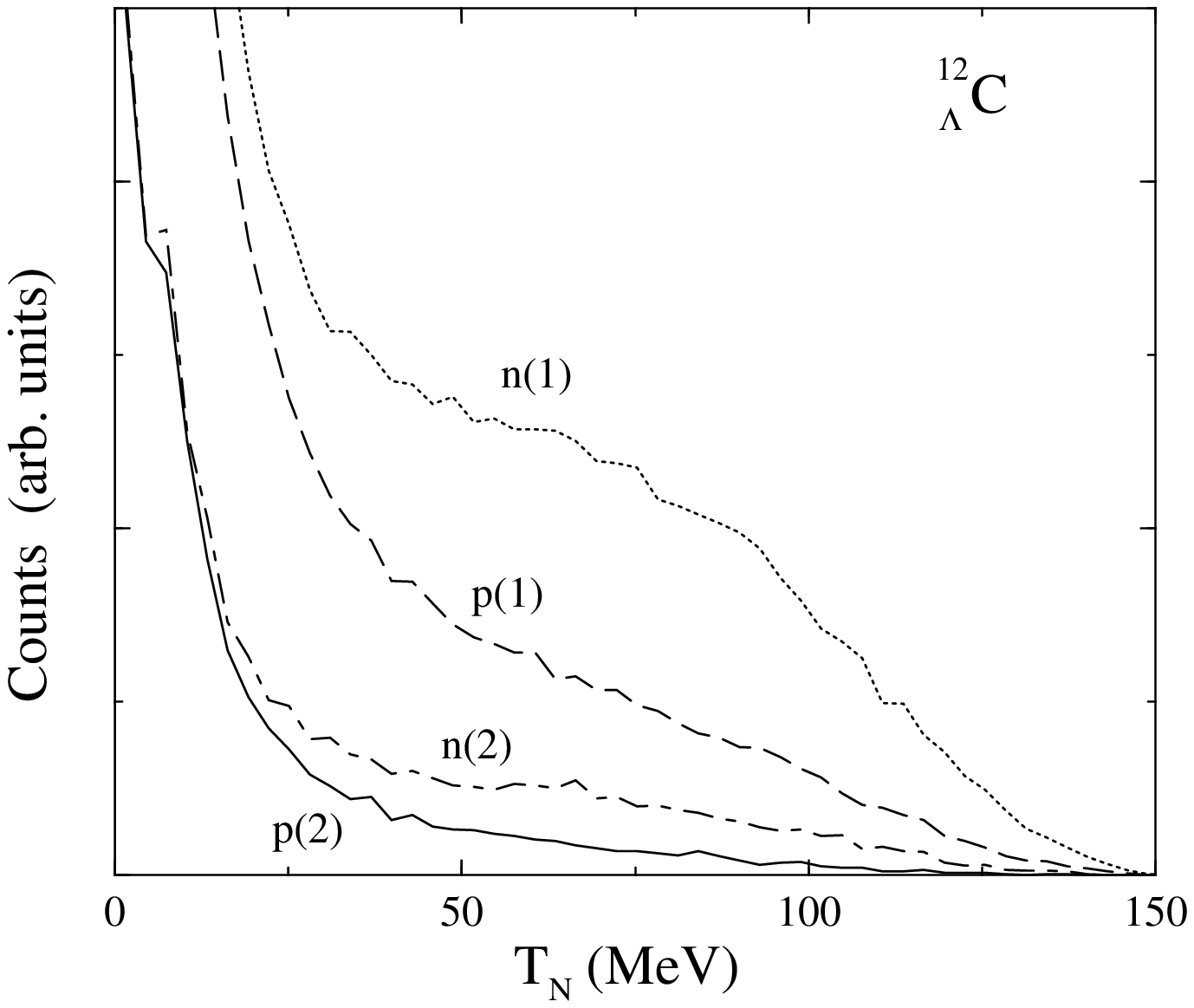}
} \caption{
 Spectra of neutrons and protons in the decay of
$^{12}_{\Lambda}$C. Dashed line: protons from the 1$N$-induced
mechanism. Dotted line: neutrons from the 1$N$-induced mechanism.
Solid line: protons from the 2$N$-induced mechanism. Dash-dotted
line: neutrons from the 2$N$-induced mechanism. The results have
been obtained for a value $\Gamma_n/\Gamma_p = 1$. }
        \label{fig:fig6}
\end{figure}
\begin{figure}[htb]
\centerline{
     \includegraphics[width=0.7\textwidth]{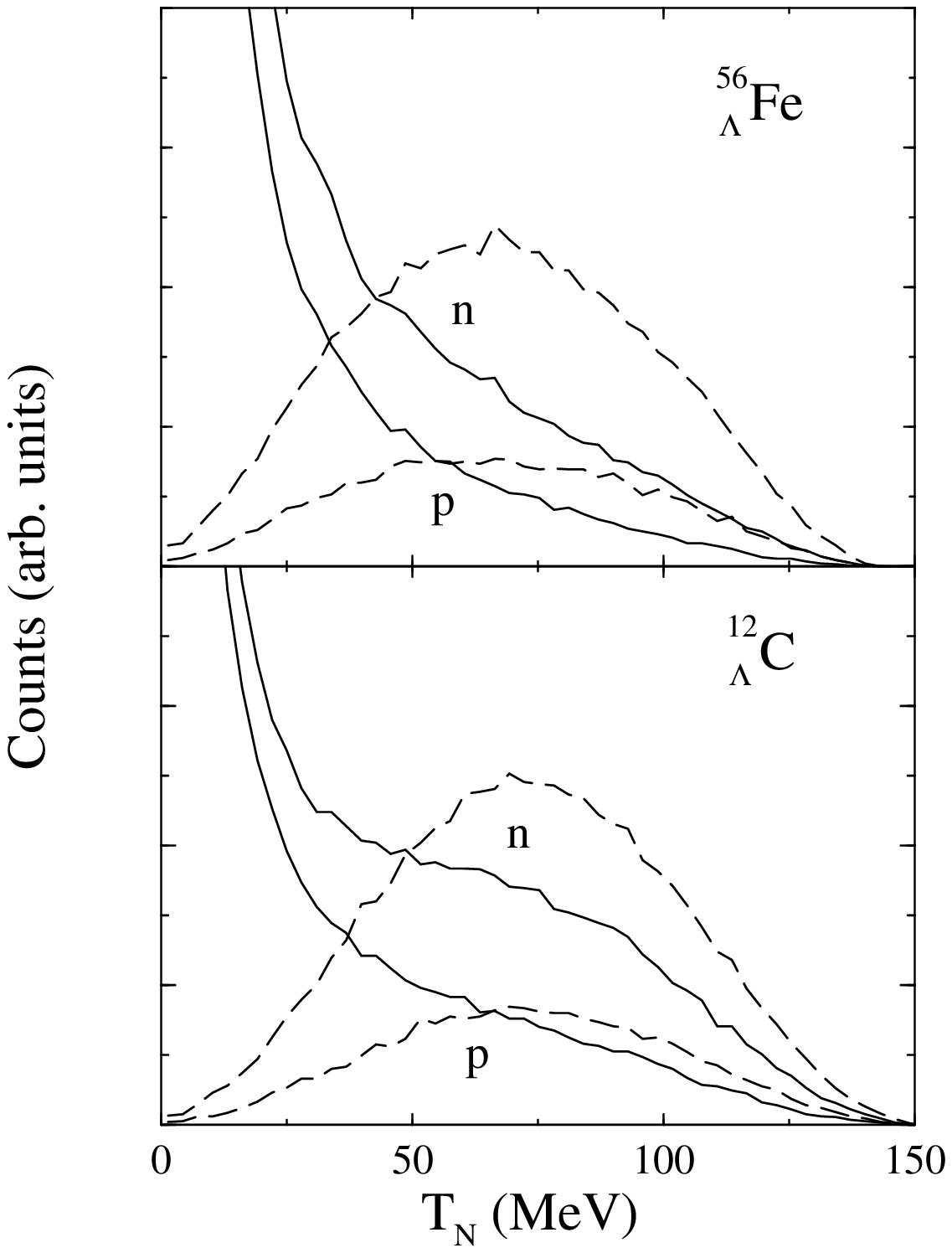}
}
      \caption{
 Effect of the final state interactions in the spectrum
of nucleons emitted in the 1$N$-induced decay of $^{56}_\Lambda$Fe
and $^{12}_\Lambda$C. Dashed line: results without FSI's. Solid
line: results including FSI's. The results have been obtained for
a value $\Gamma_n/\Gamma_p=1$. }
        \label{fig:fig7}
\end{figure}
\begin{figure}[htb]
\centerline{
     \includegraphics[width=0.8\textwidth]{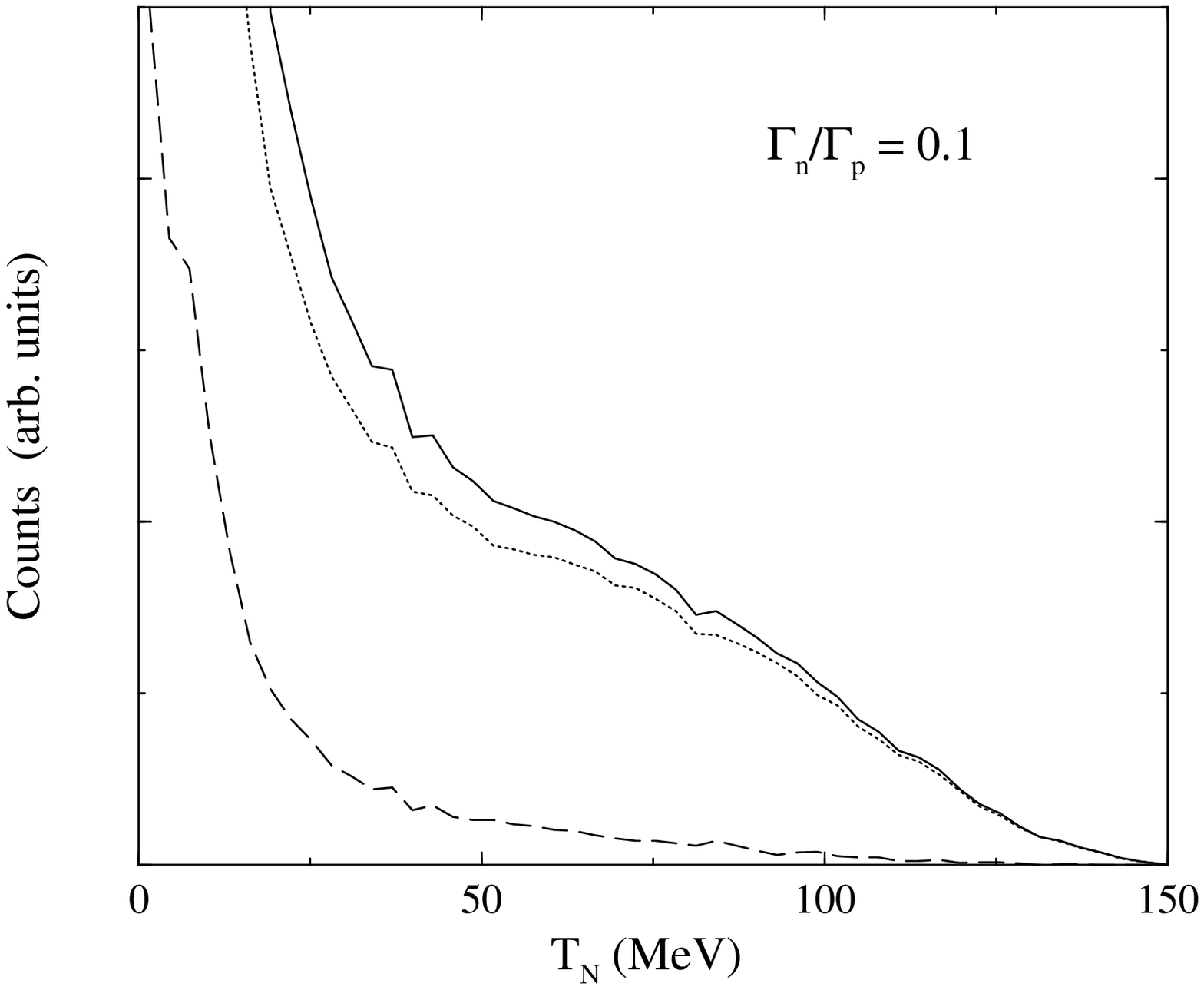}
}
      \caption{
 Proton spectrum in the decay of $^{12}_\Lambda$C,
obtained for a value $\Gamma_n/\Gamma_p = 0.1$. Dotted line:
$1N$-induced mechanism. Dashed line: $2N$-induced mechanism. Solid
line: total.
 }
        \label{fig:fig8}
\end{figure}
\addtocounter{figure}{2}
\begin{figure}[htb]
\centerline{
     \includegraphics[width=0.5\textwidth]{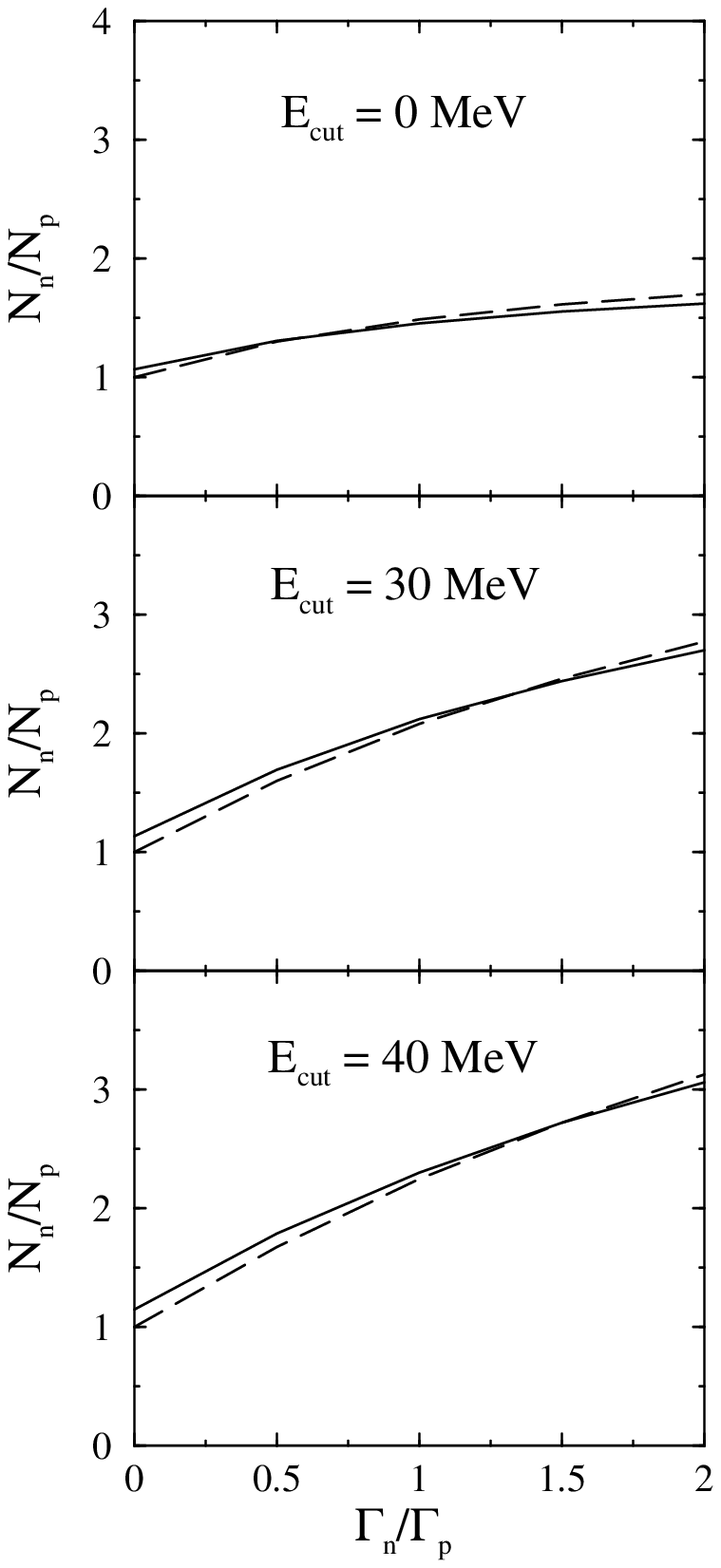}
}
      \caption{
 $N_n /N_p$ as a function of $\Gamma_n / \Gamma_p$
in the decay of $^{12}_\Lambda$C including final state interaction
effects and applying energy cuts of 0, 30, and 40 MeV. Dashed
line: $1N$-induced mechanism. Solid line: $(1N + 2N)$-induced
mechanisms.
 }
        \label{fig:fig11}
\end{figure}
\begin{figure}[htb]
\centerline{
     \includegraphics[width=0.5\textwidth]{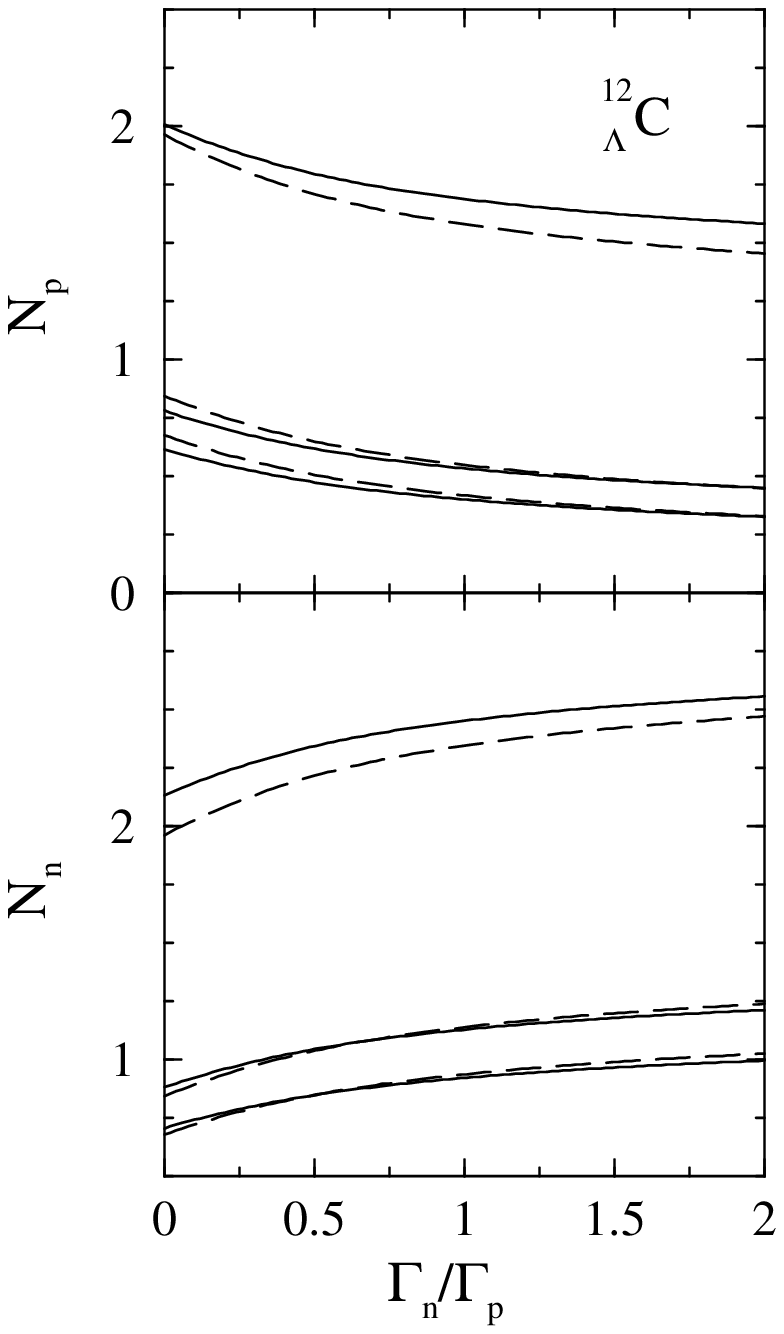}
}
      \caption{
 Number of protons $(N_p)$ and neutrons $(N_n)$  per $\Lambda$ decay
event in $^{12}_\Lambda$C as functions of $\Gamma_n / \Gamma_p$.
Dashed lines: $1N$-induced mechanism. Solid lines: $(1N +
2N)$-induced mechanisms. Final state interactions are considered
and, from top to bottom, the results include energy cuts of 0, 30,
and 40 MeV, respectively.
 }
        \label{fig:fig12}
\end{figure}
\begin{figure}[htb]
\centerline{
     \includegraphics[width=0.5\textwidth]{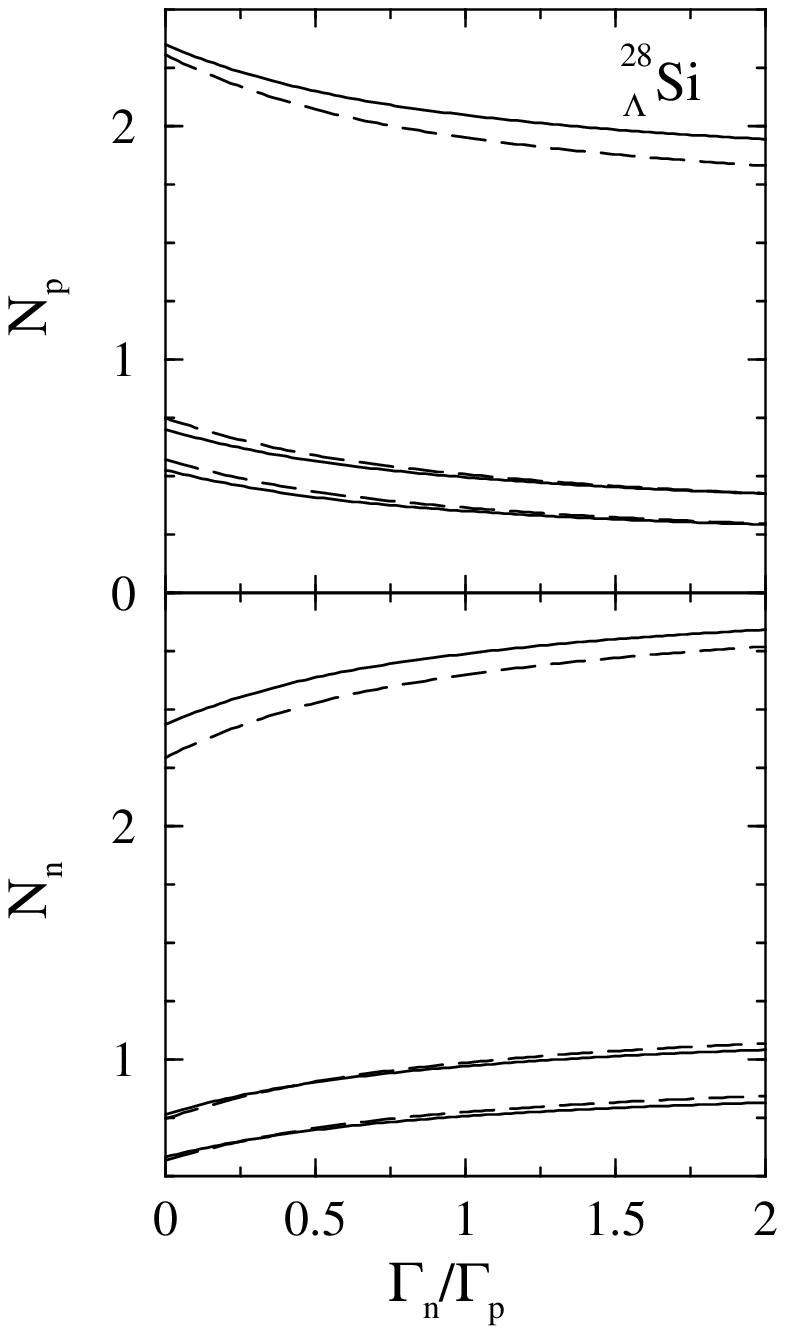}
}
      \caption{
Same as Fig.~\ref{fig:fig12} but for $^{28}_\Lambda$Si.
 }
        \label{fig:fig13}
\end{figure}
\begin{figure}[htb]
\centerline{
     \includegraphics[width=0.5\textwidth]{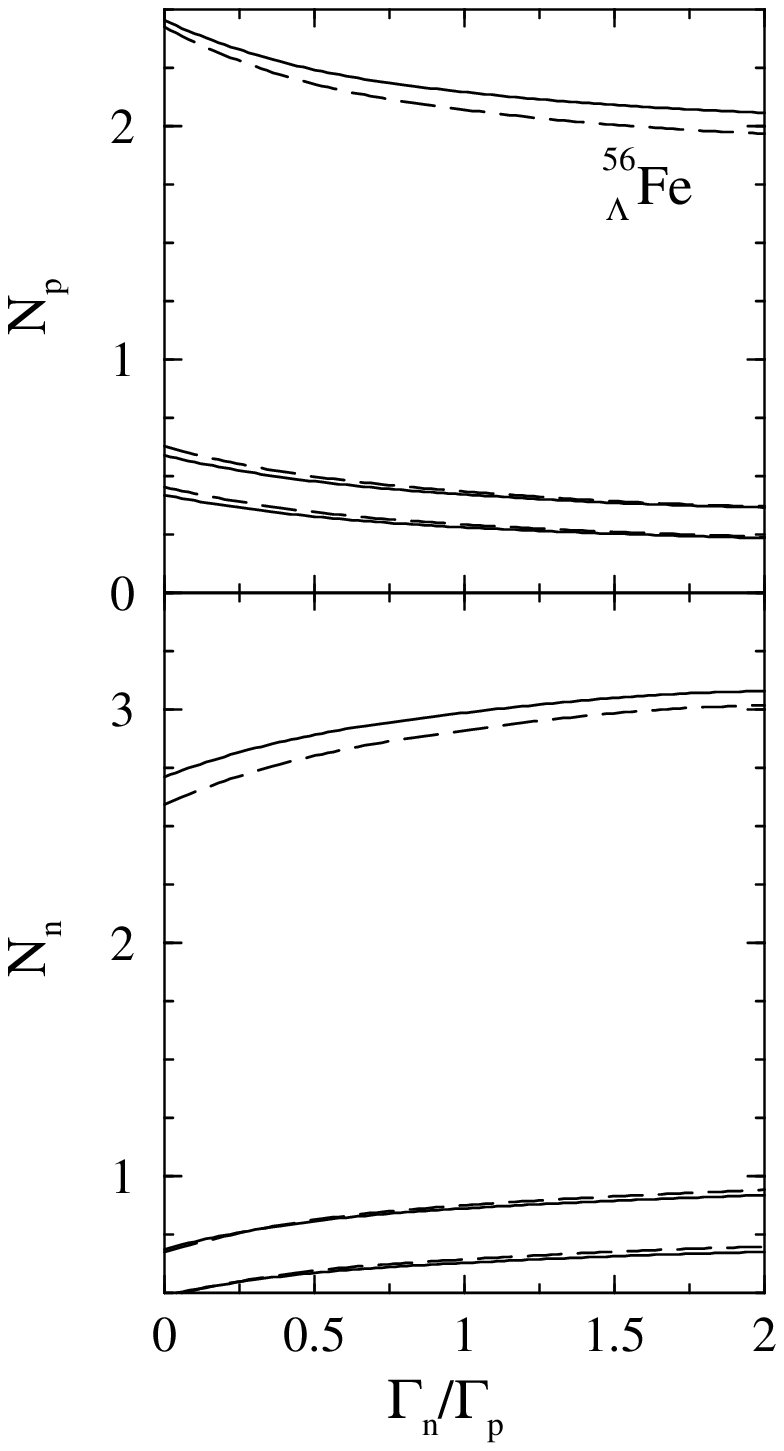}
}
      \caption{
Same as Fig.~\ref{fig:fig12} but for $^{56}_\Lambda$Fe.
 }
        \label{fig:fig14}
\end{figure}
\begin{figure}[htb]
\centerline{
     \includegraphics[width=0.8\textwidth]{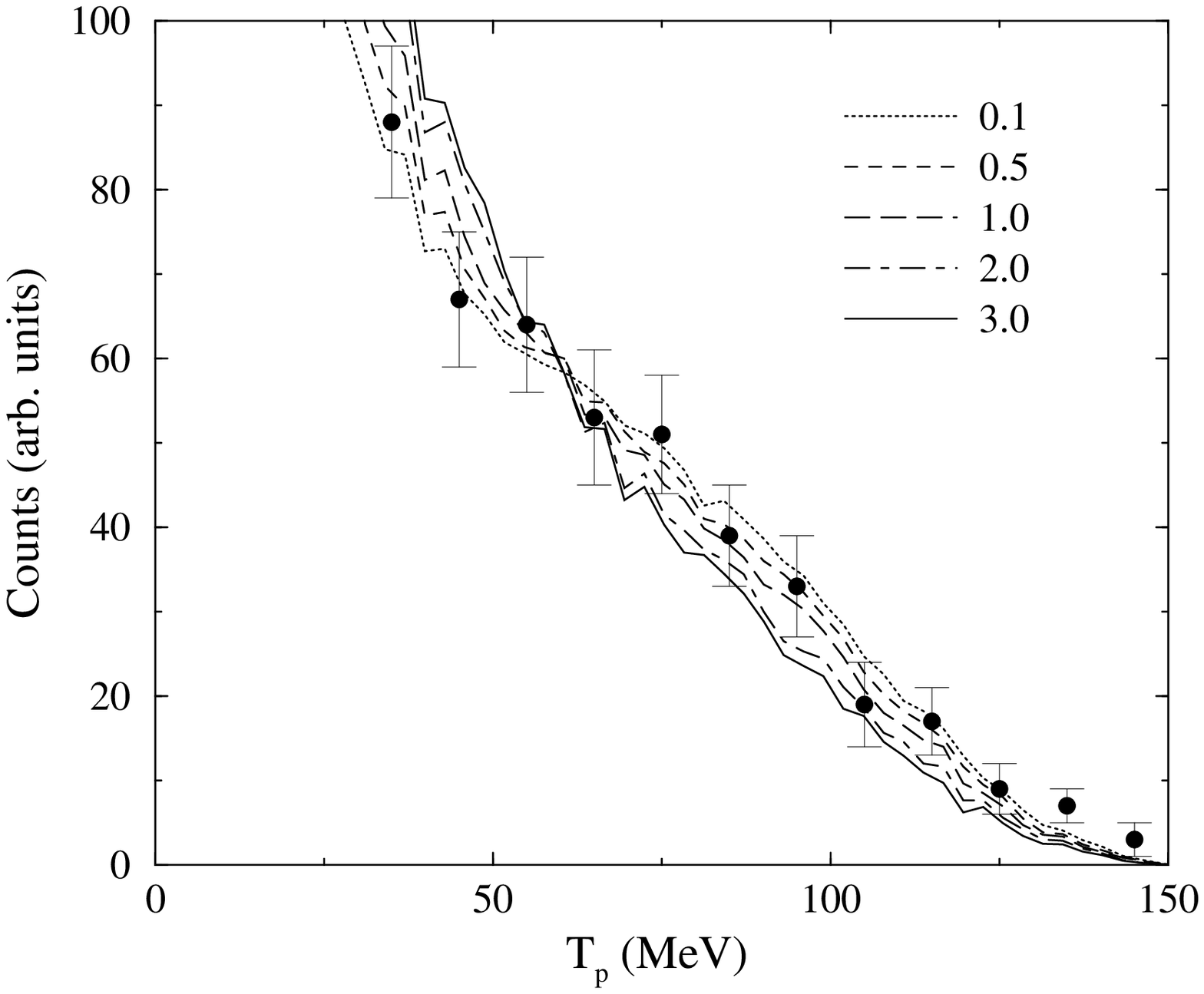}
}
      \caption{
Proton spectrum for different values of $\Gamma_n/\Gamma_p$ in the
decay of $^{12}_\Lambda$C. The experimental points are taken from
Ref.~[4].
 }
        \label{fig:fig15}
\end{figure}
\begin{figure}[htb]
\centerline{
     \includegraphics[width=0.8\textwidth]{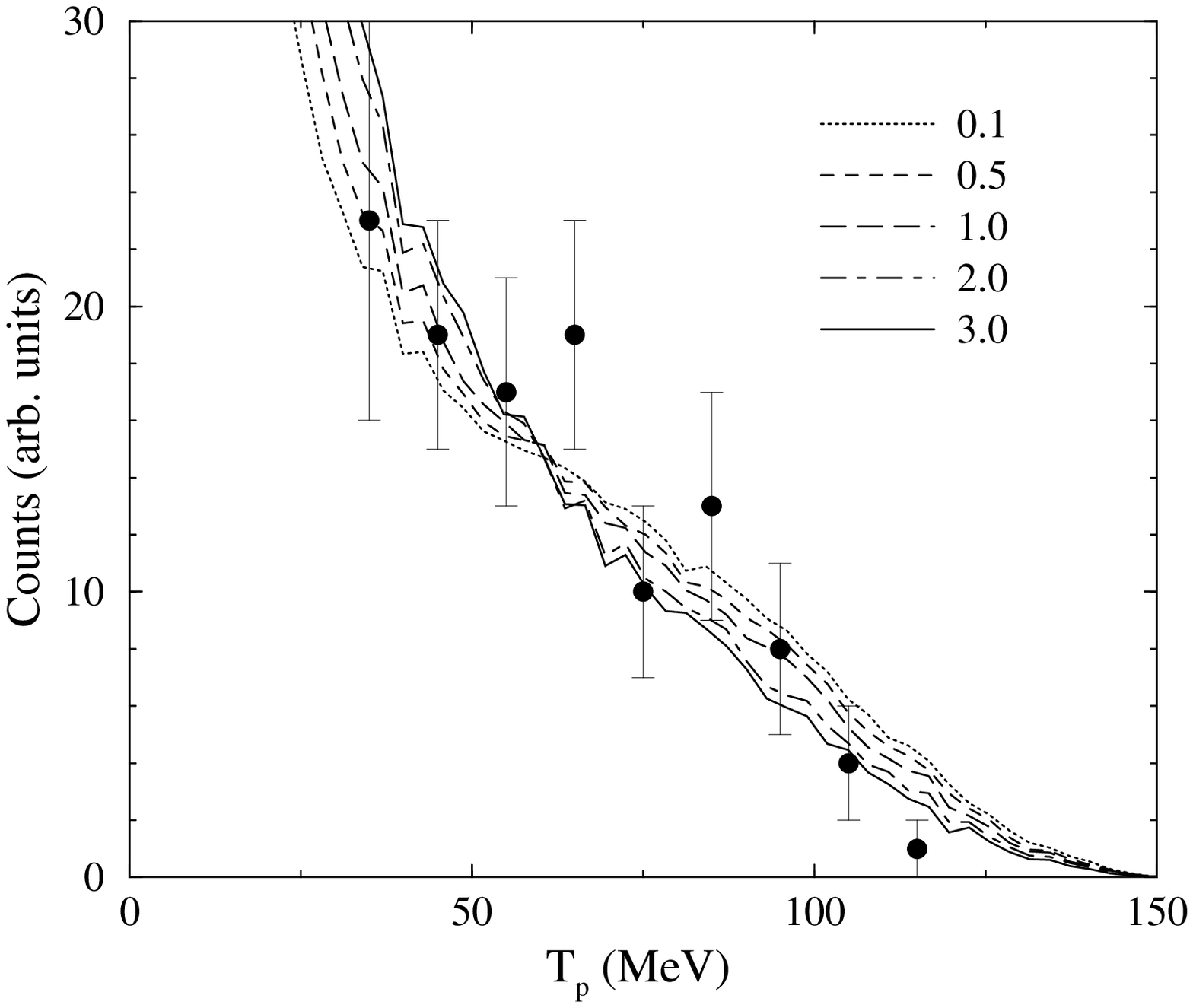}
}
      \caption{
Same as Fig.~\ref{fig:fig15} but comparing with the experimental
points of Ref.~[5].
 }
        \label{fig:fig16}
\end{figure}

\end{document}